\documentclass[twocolumn]{aastex6}
\graphicspath{{/Users/alberto/ngc5765/}}
\usepackage{comment}
\usepackage{multirow}
\usepackage{array}
\usepackage{amsmath}
\usepackage{xspace}
\usepackage{gensymb}
\newcommand{\chandra}{\textit{Chandra}\xspace}
\newcommand{\swiftbat}{\textit{Swift}-BAT\xspace}
\newcommand{\nustar}{\textit{NuSTAR}\xspace}
\newcommand{\nh}{$N_{\rm H}$\xspace}
\newcommand{\nhcgs}{cm$^{-2}$\xspace}
\newcommand{\fluxcgs}{erg cm$^{-2}$ s$^{-1}$\xspace}
\newcommand{\lumcgs}{erg s$^{-1}$\xspace}
\begin{document}
\title{Measuring the Obscuring Column of a Disk Megamaser AGN in a Nearby Merger}
\author{A. Masini\altaffilmark{1}, A. Comastri\altaffilmark{2}, R.~C.~Hickox\altaffilmark{1}, M. Koss\altaffilmark{3}, F. Civano\altaffilmark{4}, M. Brightman\altaffilmark{5}, M. Brusa\altaffilmark{6,2}, G. Lanzuisi\altaffilmark{2}}

\altaffiltext{1}{Department of Physics and Astronomy, Dartmouth College, 6127 Wilder Laboratory, Hanover, NH 03755, USA}
\altaffiltext{2}{INAF-Osservatorio di Astrofisica e Scienza dello Spazio, via Gobetti 93/3, 40129 Bologna, Italy}
\altaffiltext{3}{Eureka Scientific, 2452 Delmer Street Suite 100, Oakland, CA 94602-3017, USA}
\altaffiltext{4}{Harvard-Smithsonian Center for Astrophysics, 60 Garden Street, Cambridge, MA 02138, USA}
\altaffiltext{5}{Cahill Center for Astrophysics, California Institute of Technology, 1216 East California Boulevard, Pasadena, CA 91125, USA}
\altaffiltext{6}{Dipartimento di Fisica e Astronomia (DIFA), Universit\`a  di Bologna,  via Gobetti 93/2, 40129 Bologna, Italy}
%\altaffiltext{10}{Ambizione fellow}

\begin{abstract}
Active Galactic Nuclei (AGNs) hosting disk water megamasers are well known to be obscured by large amounts of gas, likely due to the presence along the line of sight of an almost edge-on disky structure orbiting the supermassive black hole. Correcting for the high obscuration is crucial to infer parameters intrinsic to the source, like its luminosity. 
%Since obscuration affects mainly the soft X-ray band, a broadband X-ray spectral analysis extending to the hard X-rays allows a more robust characterization of the source spectral shape.
\par We present a broadband X-ray spectral analysis of a water megamaser AGN in an early merger (NGC 5765B), combining \chandra and \nustar data. NGC 5765B is highly Compton-thick and reflection-dominated, following the general trend among disk megamasers.
Combining the exquisite black hole mass from masers with our X-ray spectroscopy, the Eddington ratio of the megamaser is estimated to be in the $2-14\%$ range, and its robustness is confirmed through SED fitting.
%The megamaser galaxy is in interaction with a nearby companion (likely active as well) and we constrain its Eddington ratio, but we argue that the early stage of the merger and the flickering nature of the accretion rate prevent any conclusive link between the AGN activity and the merger state to be assessed.

\end{abstract}

\section{Introduction} \label{sec:intro}
It is today widely accepted that supermassive black holes (SMBHs), with masses ranging between $M_{\rm BH}\sim 10^6 - 10^9 M_{\odot}$, are likely ubiquitous in the center of galaxies  \citep[e.g.,][]{kormendyrichstone95}. Such large masses are assembled over cosmic time through direct accretion of matter, and presumably through mergers of smaller black holes \citep[e.g.,][]{abbott16}. When SMBHs grow through accretion, they shine across the electromagnetic spectrum \citep[e.g.,][]{elvis94}, often dominating the emission from their host galaxy, and they are called active galactic nuclei (AGNs).
The majority of AGNs are obscured by some amount of matter, manifesting itself as a sharp flux decrease in the soft X-ray band due to photoelectric absorption \citep[e.g,][]{tozzi06,merloni14}. When the optical depth is larger than unity, the column density is larger than the inverse of the Thomson cross section (\nh $> \sigma^{-1}_T \gtrsim 1.5 \times 10^{24}$ \nhcgs) and the source is called Compton-thick (CT).
In the last decade, a small subset of AGNs -- called disk water megamasers -- was discovered to show the signature of a sub-pc scale, rotating dusty disk (or ring) in close proximity to the SMBH \citep{miyoshi95}. Such disks were discovered thanks to their peculiar water maser emission at $\sim 22$ GHz, which allows a measurement of the mass of the central SMBH with exquisite precision, and to probe the very surroundings of the AGN \citep[e.g.,][]{kuo11}. In particular, the large majority \citep[$\sim 80\%$,][]{greenhill08, castangia13,masini16} of disk water megamasers were found to show CT levels of absorption in the X-ray band. Since these disks can be detected only when almost edge-on to the observer, the large amount of obscuration suggests a natural link between the disk and the gas obscuring the AGN \citep{masini16}.
\par Recently, a new disk water megamaser was discovered in a galaxy undergoing a merger, namely NGC 5765 \citep{gao16}. The role of mergers in triggering AGN activity has been increasingly investigated both observationally \citep{koss10,cisternas11,treister12, hickox14, ricci17} and theoretically \citep{dimatteo05, hopkins06}. While AGNs in late mergers\footnote{We define early and late mergers following the definition of \citet{stierwalt13} and \citet{ricci17}; galaxies in late merging stages are separated by $\lesssim 3$ kpc \citep{koss18}.} are generally obscured by some amount of matter \citep[and references therein]{koss16a, ricci17}, as expected if the merger is driving large amounts of gas and dust towards the center of the interacting galaxies, at larger separations the situation is different. During the early stage of the merger, there is probably not enough time to make the gas lose enough angular momentum to accrete on the AGN, and the fraction of heavily obscured CT AGNs is consistent with the one measured for local, isolated galaxies from \swiftbat, where secular processes dominate the triggering of AGNs \citep{koss11,ricci15}.
\par
\begin{figure}
\plotone{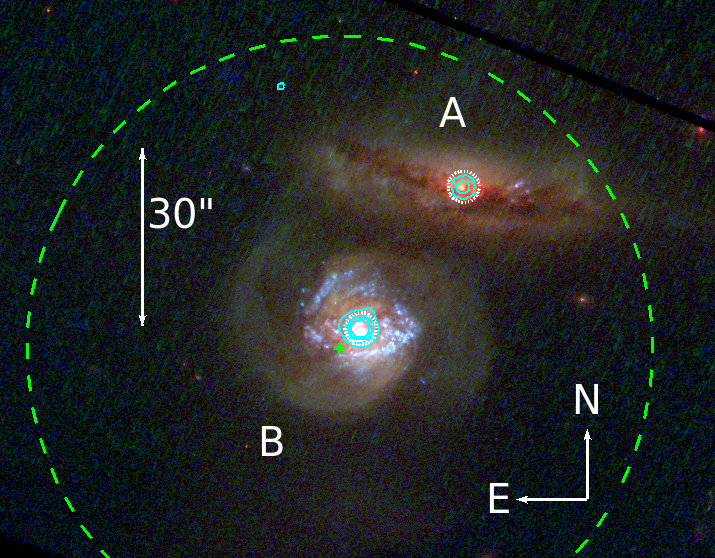}
\caption{\textit{Hubble Space Telescope} (HST) composite optical image of the merging system NGC 5765A/B, with F336W as blue, F438W as green, and F814W as red. Component A is on the top and B (the megamaser) on the bottom, as labelled. The projected separation between the two nuclei detected by \chandra is $\sim23''$ ($\sim 13$ kpc). The \chandra full band contours are superimposed in cyan, and they are comparable in size to the \chandra extraction regions (dashed white circles). The \nustar extraction region, centered almost on component B (green cross), is labelled by a green dashed circle. North is on the top, East is on the left. \label{fig:ngc5765_hstcomp_chacont}}
\end{figure}

 The NGC 5765 system is an interesting one to explore any potential observational link between the ongoing merger and the AGN activity, and/or its obscuration. It hosts a dual AGN (component A and B from here on; component B hosts the disk water megamaser), separated by a projected distance of $\sim23"$ ($\sim 13$ kpc, at $D =126.3$ Mpc; scale of $\sim 0.612$ kpc/arcsec; see Figure \ref{fig:ngc5765_hstcomp_chacont}), and can be considered as an early merger. The AGN nature of the pair is suggested by a \chandra detection in 2016, as part of a program on early dual AGNs \citep{koss12}, of both galaxies coincident with their optical nuclei in a snapshot of $\sim15$ ks, and supported by the optical emission line ratios: the SDSS optical spectrum of the component B shows narrow emission lines typical of Seyfert 2 nuclei \citep{shirazibrinchmann12}, while its companion shows more composite-like line ratios. Furthermore, the \chandra spectrum of NGC 5765B shows signature of extreme absorption (like a prominent Fe $K\alpha$ line), in agreement with the well-known recurrence of CT AGN in disk water megamasers discussed above. To help overcome the high obscuration affecting the soft X-ray band, we proposed and obtained a 50 ks observation with \nustar, the first orbiting hard X-ray ($3-79$ keV band) telescope with focusing optics \citep{harrison13}. \nustar focuses hard X-rays onto two almost identical focal plane modules (namely FPMA and FPMB), with a factor of $\sim 100$ improvement in sensitivity with respect to coded mask instruments, and opened a new window on the hard X-ray spectroscopy of both obscured and unobscured AGNs.

%How AGNs are triggered is still a debated topic with conflicting results \citep[e.g.,][]{alexanderhickox12}. Every physical mechanisms involved must be able to funnel large amount of gas in the close vicinity of the SMBH, dissipating $\sim 99\%$ of its angular momentum \citep{shlosman90}. The mechanisms often involved are the so-called secular evolution, i.e. gravitational instabilities due to motion of gas and stars in the galaxy, and wet, major mergers of galaxies \citep[e.g.,][]{kormendyho13}.

\par The paper is structured as follows: in Section \S \ref{sec:dataredution} we describe the data reduction, followed by the X-ray spectral analysis of the component A (Section \S \ref{sec:ngc5765a}) and component B (Section \S \ref{sec:ngc5765b}). Section \S \ref{sec:lum} focuses on the derivation of the intrinsic X-ray luminosity for NGC 5765B. Conclusions are presented in Section \S \ref{sec:conclusions}. 
%, used in Section \S \ref{sec:eddratio} to derive the Eddington ratio of the megamaser. 
\par Uncertainties are quoted at the 90\% confidence limit, unless stated otherwise. No cosmology was assumed to derive luminosities, since we use the angular diameter distance to NGC 5765B \citep[126.3 Mpc,][]{gao16} and its redshift \citep[$z=0.02754$,][]{ahn12}.

%\begin{figure*}
%\plotone{bpt.pdf}
%\caption{Optical diagnostic diagrams. Both sources display emission line ratios typical of AGNs. The red lines are from \citet{kewley06}. \label{fig:bpt}}
%\end{figure*}

\begin{table}
\caption{Details on the X-ray observations considered in this work.}             
\label{tab:log}      
\centering          
\begin{tabular}{l c c c c c c} 
\hline\hline       
\noalign{\vskip 0.5mm} 
 Telescope & ObsID & Date  & $t_{\rm exp}$ [ks] \\        
\noalign{\vskip 1mm} \hline  \noalign{\vskip 1mm}    
\chandra & 18158 & 2016 Mar 23  & 15 \\ \noalign{\vskip 0.5mm}
\nustar & 60301025002 & 2018 Jan 23  & 50 \\ \noalign{\vskip 0.5mm}
\noalign{\vskip 1mm}    
\hline
\end{tabular}
%\tablecomments{The area column refers to the total area, while the exposure time ($T_{\rm exp}$) column is the FPMA+FPMB exposure time at which the area drops to 0.01 deg$^2$. }
\end{table}

\section{Data reduction}\label{sec:dataredution}
\chandra ACIS-S public data were downloaded from the archive and reduced with the standard processing pipeline through \texttt{chandra\_repro} and \texttt{specextract} tasks inside CIAO \citep{fruscione06} v. 4.9, with CALDB v. 4.7.8. 
Both components of the merger are detected by \chandra, and source spectra were extracted from two circular $2"$ radius apertures, while the background was extracted from two annuli of $2"-4"$ inner and outer radii centered on each source, respectively. \chandra spectra were binned to a minimum of 3 counts per bin. NGC 5765B, the megamaser, is a factor of $\sim$ 8 brighter in the full $0.5-7$ keV \chandra band than its companion (and a factor of $\sim$ 4.5 in the $3-7$ keV band overlapping with \nustar), hence it is expected to  dominate the \nustar total flux.
\par \nustar data were reduced with the standard \texttt{nupipeline} task inside the \nustar Data Analysis Software (NuSTARDAS\footnote{\url{https://heasarc.gsfc.nasa.gov/docs/nustar/analysis/nustar_swguide.pdf}}) v. 1.8.0, with the CALDB version 20170817. In the \nustar observation, the centroid of the emission is $3.5"$ offset from the centroid of component B in the \chandra data, as expected. Moreover, the separation of the two components is comparable with the \nustar's PSF Full Width at Half Maximum. For this reason, we extracted only one spectrum from a circle of $40"$ radius (Figure \ref{fig:ngc5765_hstcomp_chacont}), while the background was extracted from two circles of $80"$ radius on the same chip of the source. The spectrum was grouped to a minimum of 5 counts per bin. Since the \nustar extraction radius is larger than the projected separation of the two components, we exploited the \chandra data to take into account the possible contaminating flux inside the \nustar aperture by component A. Table \ref{tab:log} summarizes the details of the X-ray observations considered in this work.
The Cash statistic \citep{cash79} was employed for the spectral fitting.

\section{Broadband X-ray Spectral Analysis}

\subsection{NGC 5765A}\label{sec:ngc5765a}

NGC 5765A is detected by \chandra with 27 net counts, with an hardness ratio $\sim 0$ (shown in blue in Figure \ref{fig:ngc5765b_spectrum}). With such a low statistics, we fit the spectrum with a simple power law with Galactic absorption (\nh $=2.97 \times 10^{20}$ \nhcgs, \citealp{kalberla05}), getting $\Gamma = 1.4 \pm 0.7$. We will take into account this value, along with the power law normalization at 1 keV ($N=3.4^{+2.3}_{-1.5} \times 10^{-6}$ photons keV$^{-1}$ cm$^{-2}$ s$^{-1}$), when fitting the joint \nustar + \chandra spectrum of the megamaser. If instead we assume a generally adopted value of 1.8 for the photon index \citep[e.g.,][]{burlon11} and adopt an obscured power law model\footnote{In the following analysis, the results are consistent within the uncertainties for all the tested models, adopting the obscured power law model with a fixed photon index $\Gamma=1.8$ and leaving the column density free to vary.}, we obtain a column density \nh $\lesssim 9\times10^{23}$ \nhcgs. A multiwavelength study of the dual AGN nature of this system is part of a forthcoming study (Koss et al in prep).

\subsection{NGC 5765B}\label{sec:ngc5765b}

\begin{figure}
\plotone{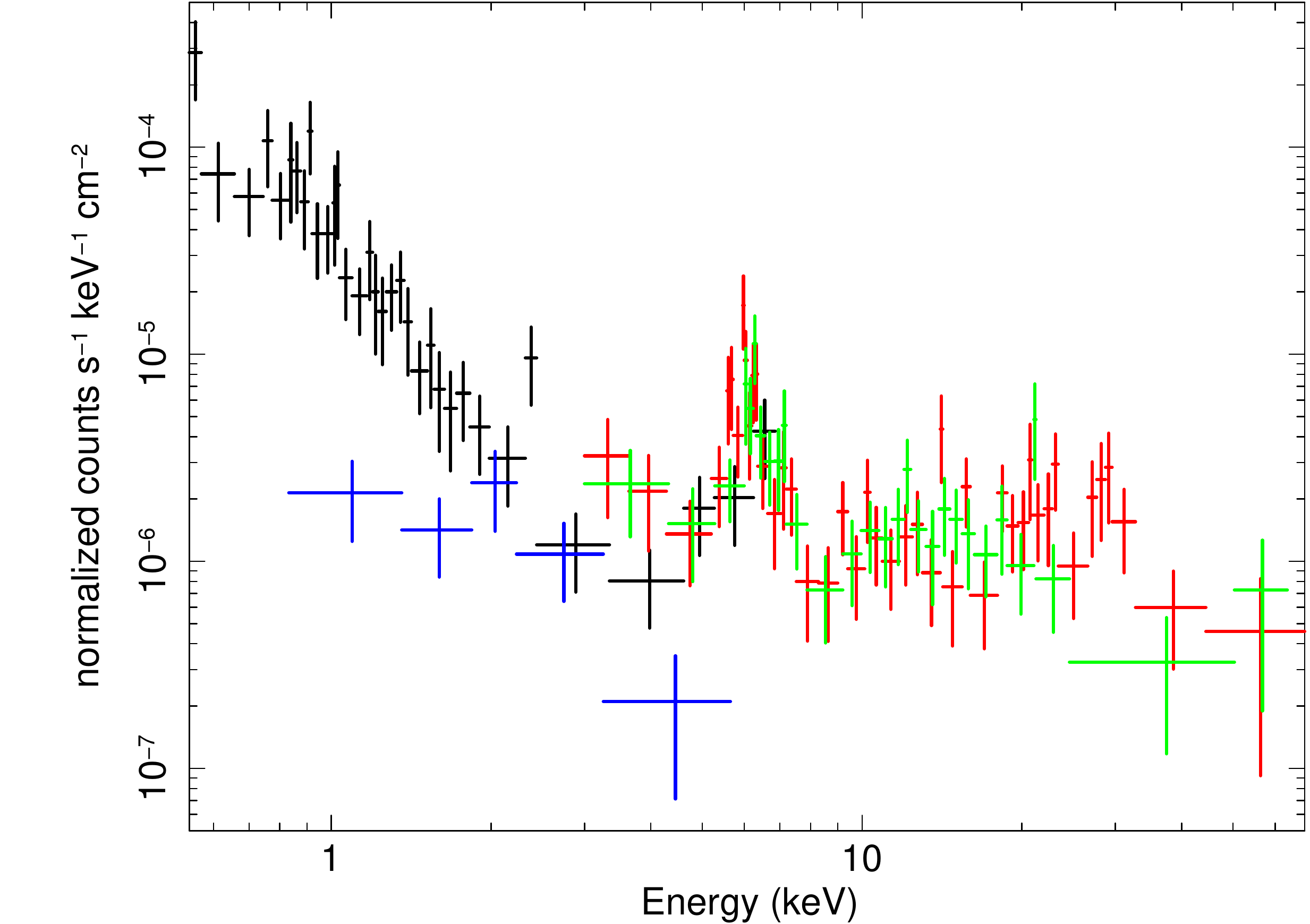}
\caption{Joint \chandra (black) and \nustar (red and green for FPMA and FPMB, respectively) X-ray spectrum of NGC 5765B, and \chandra spectrum of NGC 5765A (blue). The data have been divided by the response effective area for each channel, and have been rebinned for plotting purposes. The \nustar spectrum is likely contaminated by the contribution of NGC 5765A.\label{fig:ngc5765b_spectrum}}
\end{figure}

NGC 5765B is detected by \chandra with 197 net counts, and by \nustar with 349 and 278 (FPMA and FPMB, respectively) net counts.
The joint \nustar + \chandra spectrum of the megamaser shows a soft component below few keV, a hard X-ray flat spectral shape, and an emission feature around $\sim 6$ keV (Figure \ref{fig:ngc5765b_spectrum}). These features are more prominent when fitting the whole spectrum with a single power law, that is clearly unable to capture the spectral complexity of heavily obscured AGNs. We then build a phenomenological model, consisting of a power law (to account for the contamination of NGC 5765A in the \nustar spectrum\footnote{We force the photon index and normalization parameters to be within the 90\% confidence limits found in \S\ref{sec:ngc5765a}. Moreover, the normalization of this component is set to zero in the \chandra data.}), a soft diffuse X-ray emission from collisionally-ionized plasma (an \texttt{apec} component in XSPEC) to model the very soft emission below few keV, an absorbed, intrinsic power law  \citep[\texttt{plcabs};][]{yaqoob97}, Compton reflection\footnote{The reflection parameter is fixed to a negative value, in order to have a "pure" reflection component} \citep[a \texttt{pexrav} component;][]{pexrav95}, and two Gaussian emission lines to model two line features: the first one is directly visible in the spectrum, the second one was added after examining the residuals. In XSPEC notation, the model is the following:

\begin{equation}
\label{eq:m0}
\begin{split}
\mathrm{M0} =\overbrace{ \mathrm{constant}}^{\text{cross-normalization}}\times\overbrace{\mathrm{phabs}}^{\text{Gal. absorption}}\times \\
\times~ (\overbrace{ \mathrm{plcabs} + \mathrm{pexrav} + \mathrm{zgauss} + \mathrm{zgauss} }^{\text{nuclear emission}}~ +\\
  +\underbrace{\mathrm{zpowerlw}}_{\text{contamination}} +\underbrace{\mathrm{apec)}}_{\text{soft component}}.
\end{split}
\end{equation}

\begin{figure}
\plotone{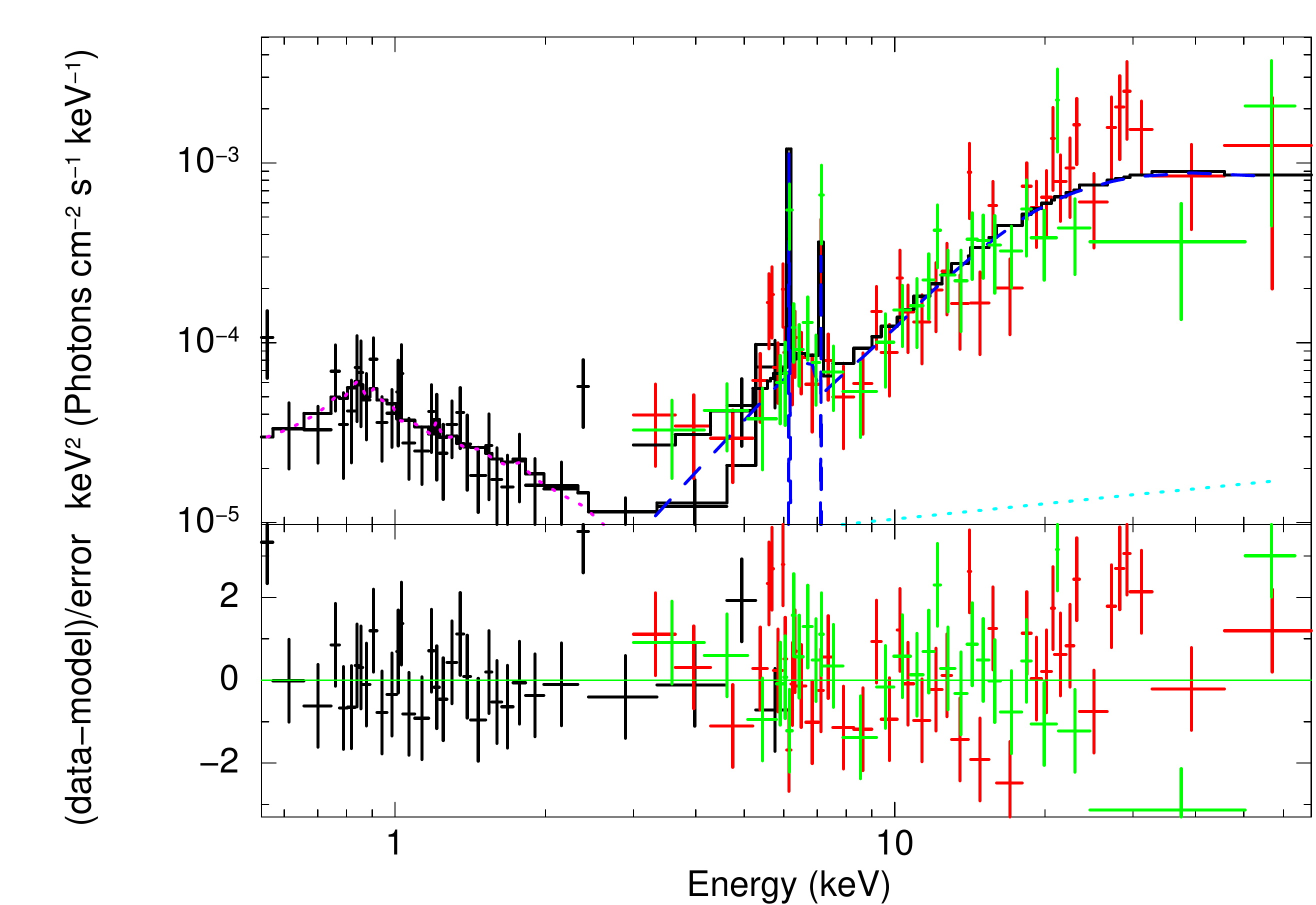}
\caption{Unfolded data and model, along with the deviation from the fit (bottom panel), for model M0. Black, red and green data are \chandra, \nustar FPMA, and FPMB, respectively. The best fit is marked by a thick black line. The different components of the best fit model are also shown. Reflection and line components are labelled by dashed blue lines, while dotted lines refer to the soft, \texttt{apec} component (magenta) and contaminant power law from NGC 5765A in the \nustar data only (cyan). \label{fig:m0}}
\end{figure}

\begin{figure}
\plotone{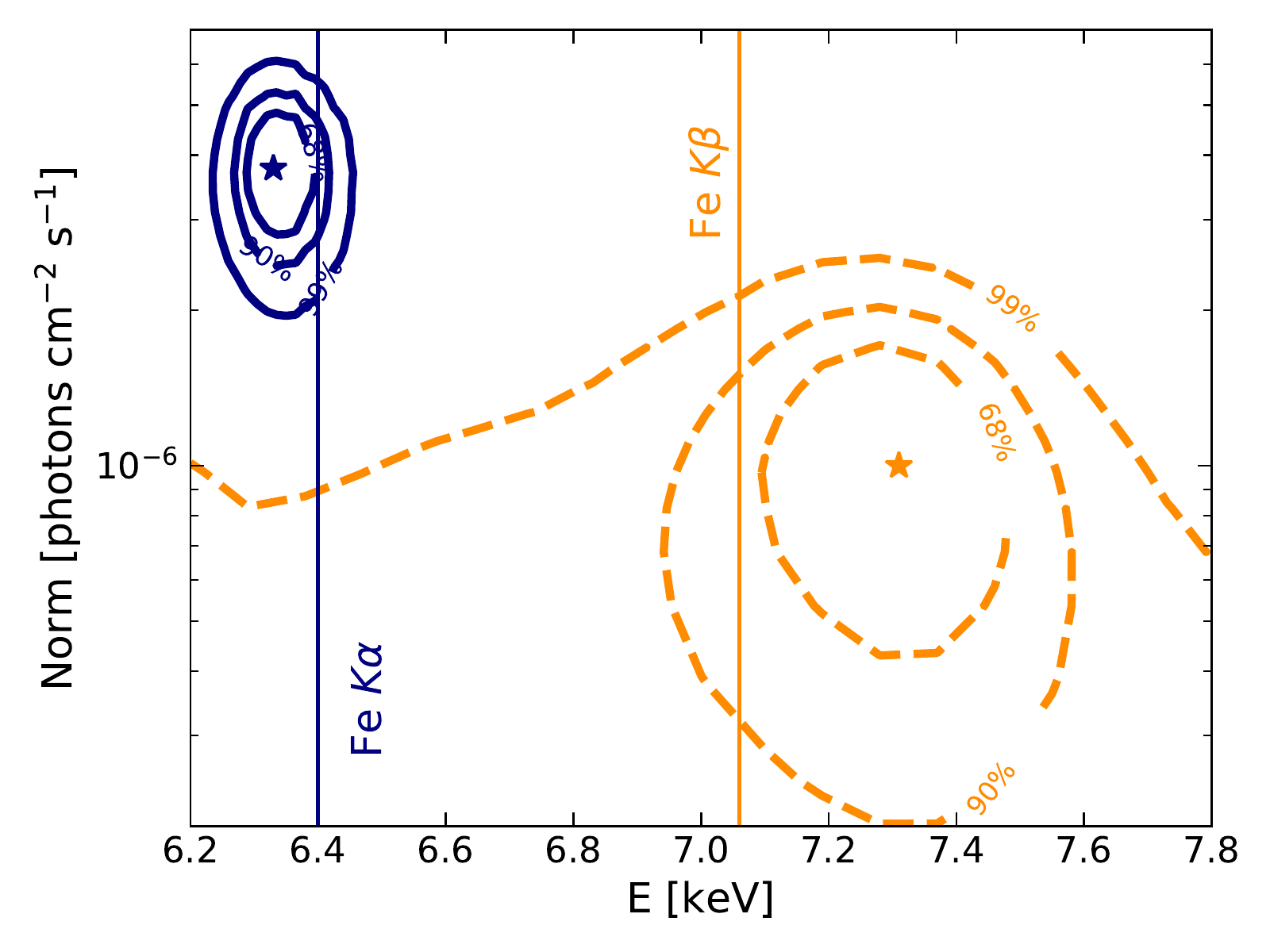}
\caption{Confidence contours ($68\%$, $90\%$, and $99\%$) for the energy and normalization of the two line features detected in the X-ray spectrum of NGC 5765B, in solid dark blue and dashed orange lines. The best fit values are marked by colored stars. The features are consistent with being Fe $K\alpha$ and Fe $K\beta$ lines (dark blue and orange vertical lines, respectively). \label{fig:lines}}
\end{figure}

A good fit is obtained (CSTAT/$\nu = 252/247$), where the reflection component dominates over the intrinsic one. The \texttt{plcabs} component is indeed not significant, the column density is unconstrained and removing it does not affect the fit. This suggests that NGC 5765B is reflection dominated. The power law modeling the contamination of NGC 5765A in the \nustar data is not significantly required by the data, but removing it makes the ACIS-S/FPMA ratio no longer consistent with unity at the 90\% confidence limit.  The centroid energies of the two line features are consistent with the simplest assumption of being Fe $K\alpha$ and Fe $K\beta$, as shown in Figure \ref{fig:lines}, and fixing the energies to those of Fe $K\alpha$ and $K\beta$ worsens the fit at the 93\% of confidence limit. Also, removing the Fe $K\beta$ line gives a worse fit at the 97\% confidence limit. The lines have large equivalent width, as reported in Table \ref{tab:results}. Finally, the thermal component has $kT=0.8^{+0.2}_{-0.3}$ keV and a subsolar abundance ($Z/Z_\odot < 0.08$), which is not uncommon and likely a consequence of a one-temperature model \citep[e.g.,][]{greenhill08}.

\subsubsection{MYTorus model}\label{sec:myt}
Given that the hard X-ray spectrum of NGC 5765B is reflection dominated and the data don't require an intrinsic power law, a \texttt{pexrav} model alone is not able to give a reliable estimate of the column density obscuring the nucleus. This extreme obscuration was expected due to the well known high incidence of CT obscuration within water disk megamasers \citep{greenhill08, castangia13, masini16}, and also on the basis of the toy model of \citet{masini16}, which relates the obscuring column density to the extension of the disk along the line of sight. Adopting the inner and outer maser disk radii from \citet{gao16}, the predicted column density is \nh $\sim 7 \times 10^{24}$ \nhcgs. 
\par In an attempt to have a robust estimate of the column density, we employ a MYTorus model \citep{murphy09}. It has been designed to fit CT AGN spectra, and has been extensively used in the literature in its default and decoupled modes. In its default mode, the model assumes a toroidal, uniform medium with a covering factor 0.5, and self-consistently computes the transmitted (namely component MYT$_{\rm Z}$ obscuring the intrinsic power law), scattered (namely MYT$_{\rm S}$) and fluorescence emission  (namely MYT$_{\rm L}$) through it. In this case, a default MYTorus model does not provide a good fit (CSTAT/$\nu = 307/251$), significantly underestimating the line emission. 
\par The components of MYTorus, usually linked to each other during the fit, can be decoupled to simulate more complex geometries \citep[see][]{yaqoob12}. For instance, allowing the scattered and fluorescence components to have a face-on inclination (i.e., inclination angle of $0\degree$, with the components labeled $MYT_{S,00}$ and $MYT_{L,00}$, respectively), while the intrinsic one is kept fixed to edge-on at an inclination angle of $90\degree$ (and labeled $MYT_{Z,90}$) we simulate a geometry in which the obscuring medium is clumpy, and most of the reflection comes from unobscured lines of sight, or from more distant material not suffering of significant nuclear attenuation. 
\par This decoupled configuration of MYTorus provides a good fit (CSTAT/$\nu = 251/250$), with the column density hitting the upper cap allowed by the model (i.e., \nh $= 10^{25}$ \nhcgs). The column density obscuring the intrinsic component and the one producing the reflection spectrum are statistically consistent ($N_{\rm H}^{Z,90} > 3 \times 10^{24}$ \nhcgs and $N_{\rm H}^{A,00} > 5 \times 10^{24}$ \nhcgs, respectively), and as such they are kept linked. We refer to this model as M1:

\begin{equation}
\label{eq:m1}
\begin{split}
\mathrm{M1} =\overbrace{ \mathrm{constant}}^{\text{cross-normalization}}\times\overbrace{\mathrm{phabs}}^{\text{Gal. absorption}}\times \\
\times~ (\overbrace{ \mathrm{zpowerlw}\times \mathrm{MYT_{Z,90}}}^{\text{absorption}} + \overbrace{\mathrm{MYT_{S,00}} + \mathrm{MYT_{L,00}} }^{\text{back-scattering}}~ +\\+\underbrace{\mathrm{zpowerlw}}_{\text{contamination}} + \underbrace{\mathrm{apec}}_{\text{soft component}}).
\end{split}
\end{equation}

\begin{figure}
\plotone{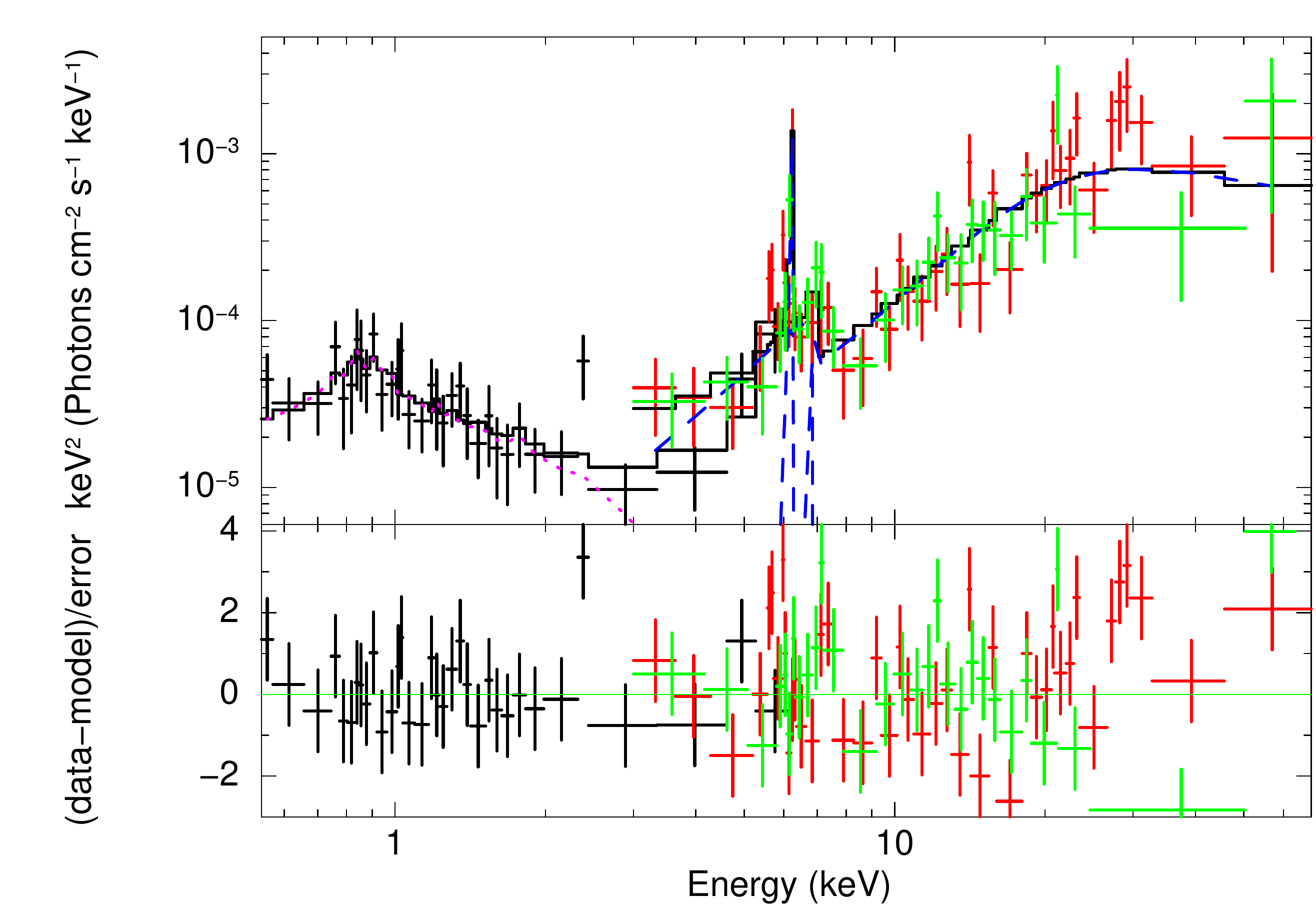}
\caption{Same of Figure \ref{fig:m0}, but for model M1. \label{fig:m1}}
\end{figure}

The best fit parameters for model M1 are shown in Table \ref{tab:results}. As can be seen, the cross-normalization constant between \chandra's ACIS-S and \nustar's FPMs is ACIS-S/FPMA$=0.6^{+0.3}_{-0.2}$. 
\par Both the M0 and M1 model show a deviation from the unitary cross-normalization constant between \chandra's ACIS-S and \nustar's FPMA ($\sim 0.6$, although model M0 is consistent with unity within the uncertainty), even taking into account the emission from NGC 5765A in the \nustar data. 
This discrepancy between the \chandra and \nustar data could be due to a column density variability (the data suggests that NGC 5765B could have transitioned from marginally CT to heavily CT after the \chandra observation), or to a large amount of reflected flux originating at large distances (larger than kpc-scale) from the nucleus, similarly to what is already seen in other CT AGNs \citep{marinucci12,arevalo14,bauer15,fabbiano17}. If this is the case, only the \nustar spectrum would include the extended hard X-ray emission due to the larger extraction region, while the 2" \chandra extraction region, corresponding to $\sim1.2$ kpc of radius, would not (see Figure \ref{fig:ngc5765b_spectrum}). 

%The \nustar spectrum seems indeed to be flatter at $3 < E < 5$ keV (see Figures \ref{fig:m0} and \ref{fig:m1}). Even if this discrepancy is just at the $1\sigma$ level, we tried to fit the X-ray spectrum with model M1, fixing the \chandra-\nustar cross-normalizations to one, but allowing the column density between the \nustar and \chandra observations to vary. In other words, we are fitting the \chandra data with a default MYTorus model, and the \nustar data with a decoupled MYTorus (plus the contamination from NGC 5765A). The column densities are linked among the different components of the models. The fit is slightly worse than the one with model M1 (CSTAT/$\nu$ = 255/250), and interestingly the column densities are significantly different: while the \nustar spectrum is heavily CT (\nh $> 5 \times 10^{24}$ \nhcgs), the \chandra spectrum is mildly CT (\nh $= 1.2^{+ 0.3}_{- 0.2} \times 10^{24}$ \nhcgs). 
%Interestingly, . Another possible explanation could be a true column density variability between the \chandra and \nustar observations.
%\par Either case, the difference in cross-normalization between the two instruments it is not sufficiently significant to be kept into consideration in the subsequent analysis.

\begin{figure}
\plotone{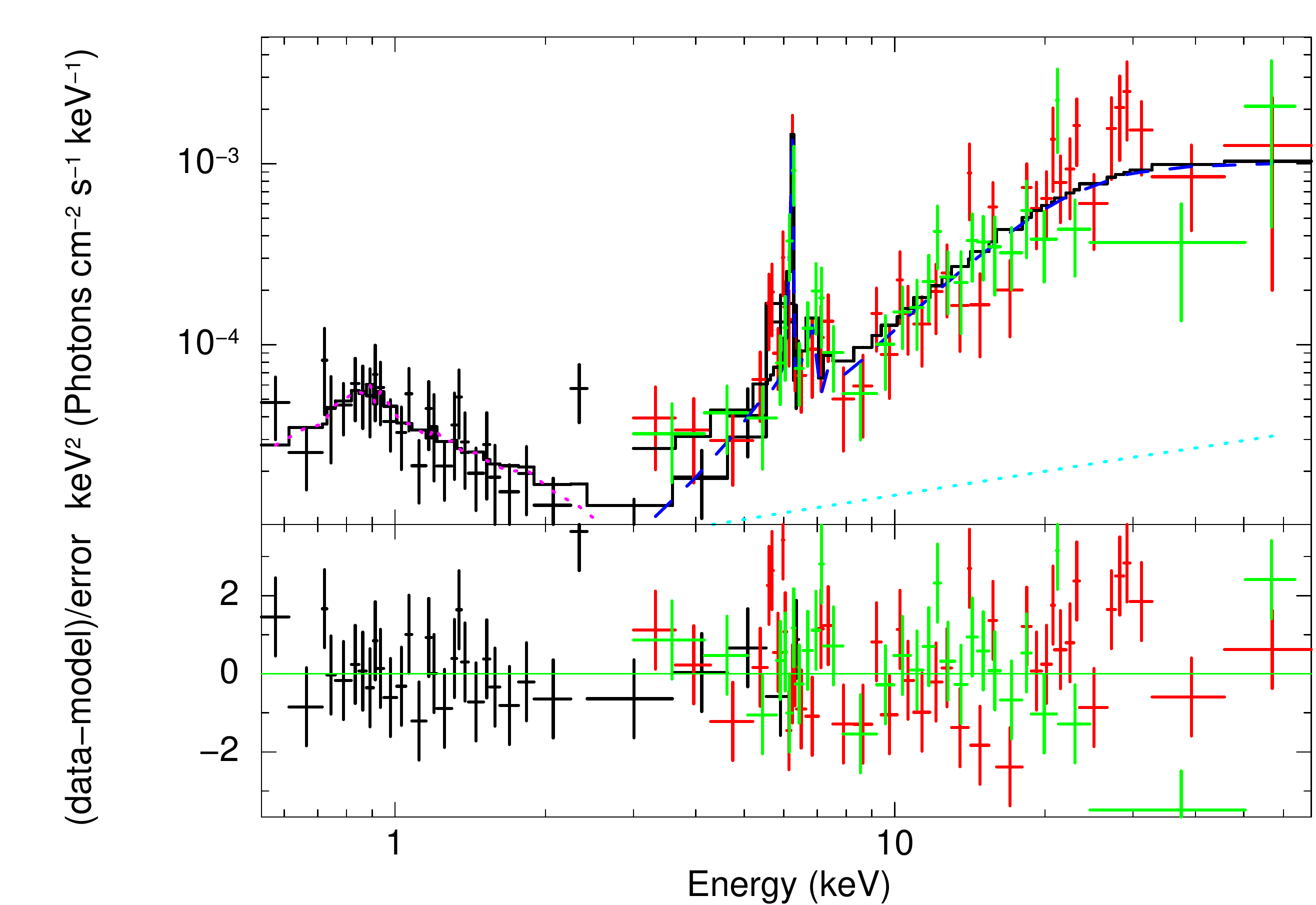}
\plotone{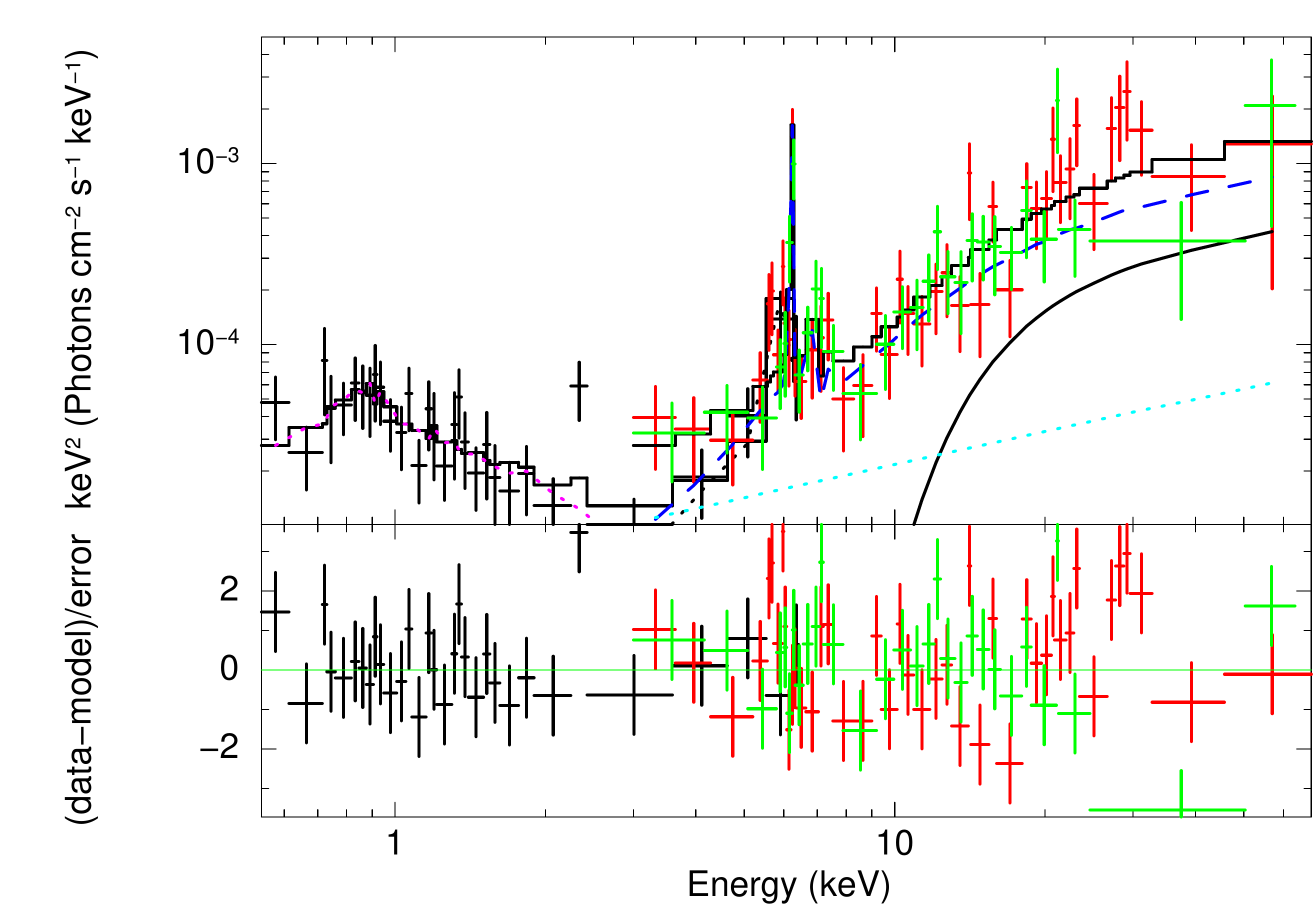}
\caption{Same of Figure \ref{fig:m0} and Figure \ref{fig:m1}, but for model M2 (top) and M3 (bottom). In model M3, the solid black line labels the absorbed intrinsic emission.\label{fig:m2-3}}
\end{figure}

\subsubsection{Borus02 model}\label{sec:borus}
As discussed in the previous Section (\S \ref{sec:myt}), the column density of the MYTorus model hits the upper cap at \nh $=10^{25}$ \nhcgs. Recently, \citet{balokovic18} published a new toroidal model, called Borus02, whose column density ranges till \nh $=10^{25.5}$ \nhcgs. Briefly, Borus02 models a uniform density sphere with polar cutouts, where the covering factor (i.e. the cosine of the half-opening angle of the torus) is a free parameter. It is important to note that the column density obscuring the line of sight component and the average column density of the torus can be different, in order to simulate the known clumpiness of the torus. In XSPEC notation, we define the following model:

\begin{equation}
\label{eq:m2}
\begin{split}
\mathrm{M2} =\overbrace{ \mathrm{constant}}^{\text{cross-normalization}}\times\overbrace{\mathrm{phabs}}^{\text{Gal. absorption}}\times \\
\times~ (\overbrace{ \mathrm{cutoffpl}\times \mathrm{zphabs}\times \mathrm{cabs}}^{\text{absorption}} + \overbrace{\mathrm{Borus02}}^{\text{reprocessing}}~ +\\+\underbrace{\mathrm{zpowerlw}}_{\text{contamination}} + \underbrace{\mathrm{apec}}_{\text{soft component}}).
\end{split}
\end{equation}

When fitting the megamaser's spectrum with model M2 (top panel of Figure \ref{fig:m2-3}), the column density along the line of sight and the average torus' one were kept linked, and the covering factor was fixed to be CF$=0.5$. The resulting fit is shown in Table \ref{tab:results}. Despite the larger dynamic range in column density, model M2 is not able to constrain it, and \nh hits the higher limit of $10^{25.5}$ \nhcgs. 
\par Finally, we also explored a different configuration of Borus02, imposing a disk-like covering factor (fixing the CF$=0.1$) and allowing the column densities (along the line of sight and the torus average one) to be different. We refer to this model as model M3. Model M3 is statistically indistinguishable from model M2, although we prefer its configuration because it simulates a clumpy, disk-like torus. This choice is motivated by the presence of the edge-on megamaser disk, which we are implicitly assuming to be at least linked to the obscurer itself. Results from model M3, summarized in Table \ref{tab:results}, suggest that the both the average torus column density (the Borus02 one) and the one along the line of sight are consistently above the CT level. Interestingly, adopting either model M2 and M3, the ratio of the ACIS-S/FPM instruments gets consistent with unity.

\begin{deluxetable*}{lcccc}
\tablecaption{Results of X-ray spectroscopy of NGC 5765B\label{tab:results}.}
\tablehead{
\colhead{Parameter} & \colhead{M0 } & \colhead{M1} & \colhead{M2 } & \colhead{ M3} \\
 & \colhead{Phenomenological} & \colhead{MYTorus decoupled} & \colhead{Borus02} & \colhead{Disky Borus02}
}
%\colnumbers
\startdata
CSTAT/$\nu$                                                 & 252/247                               & 251/250                               & 250/249                               & 251/249                               \\
ACIS-S/FPMA                                               & $0.6^{+0.5}_{-0.2}$                   & $0.6^{+0.3}_{-0.2}$                 & $0.8^{+0.5}_{-0.2}$                & $0.8^{+0.5}_{-0.3}$                \\
FPMB/FPMA                                                 & $0.9^{+0.2}_{-0.1}$                & $0.9^{+0.2}_{-0.1}$                & $0.9^{+0.2}_{-0.1}$                & $0.9^{+0.2}_{-0.1}$                \\
\noalign{\vskip 0.5mm} \hline \noalign{\vskip 0.5mm}
$kT$ [keV] & $0.8^{+0.2}_{-0.2}$ & $0.9^{+0.1}_{-0.2}$ & $0.9^{+0.1}_{-0.2}$ & $0.9^{+0.1}_{-0.2}$ \\
$Z/Z_{\odot}$ & $2.0^{+5.5}_{-l} \times 10^{-2}$ & $3.5^{+6.8}_{-3.1} \times 10^{-2}$ & $2.8^{+4.9}_{-1.5} \times 10^{-2}$ & $2.9^{+5.6}_{-2.6} \times 10^{-2}$ \\
Norm$_{\rm Apec}$ [erg cm$^{-2}$ s$^{-1}$ keV$^{-1}$] & $4.4^{+4.0}_{-2.2} \times 10^{-4}$ & $3.8^{+3.4}_{-1.8} \times 10^{-4}$ & $2.9^{+2.3}_{-1.2} \times 10^{-4}$ & $2.7^{+2.3}_{-1.2} \times 10^{-4}$ \\
\noalign{\vskip 0.5mm} \hline \noalign{\vskip 0.5mm}
$\Gamma$                                                  & $1.6^{+0.2}_{-0.3}$                & $2.3^{+0.2}_{-0.2}$                & $1.8^{+0.1}_{-0.1}$                & $1.6^{+0.3}_{-l}$                   \\
Norm [erg cm$^{-2}$ s$^{-1}$ keV$^{-1}$]             & $3.2^{+3.1}_{-1.1} \times 10^{-4}$ & $5.2^{+3.3}_{-2.4} \times 10^{-3}$ & $4.5^{+3.7}_{-0.4} \times 10^{-4}$ & $1.2^{+1.5}_{-0.6} \times 10^{-3}$ \\
\nh [$\times 10^{24}$ \nhcgs]                                      & $-$                                   & $10^{+u}_{-5.1}$                      & $18^{+u}_{-15}$                   & $3.4^{+2.8}_{-0.9}$                   \\
$N_{\rm H, tor}$ [$\times 10^{24}$ \nhcgs]                          & $-$                                   & $-$                                   & $=N_{\rm H}$                                   & $10.0^{+u}_{-9.7}$        \\
%$\Gamma_{\rm Soft}$                                       & $3.3^{+0.3}_{-0.3}$                & $3.3^{+0.4}_{-0.3}$                & $3.2^{+0.3}_{-0.4}$                & $3.2^{+0.3}_{-0.3}$                \\
%Norm$_{\rm Soft}$ [erg cm$^{-2}$ s$^{-1}$ keV$^{-1}$] & $8.7^{+6.3}_{-4.0} \times 10^{-5}$ & $9.4^{+5.3}_{-4.0} \times 10^{-5}$ & $6.6^{+4.6}_{-2.8} \times 10^{-5}$ & $6.3^{+4.4}_{-2.5} \times 10^{-5}$ \\
$E_{\rm K\alpha}$  [keV]                            &  $6.33^{+0.07}_{-0.05}$               &  $-$  &   $-$   & $-$       \\  
$EW_{\rm K\alpha}$  [keV]                            &  $1.8^{+0.5}_{-0.5}$               &  $-$  &   $-$   & $-$      \\  
$E_{\rm K\beta}$  [keV]                            &  $7.31^{+0.19}_{-0.25}$               &  $-$  &   $-$   & $-$         \\  
$EW_{\rm K\beta}$  [keV]                            &  $0.7^{+0.5}_{-0.5}$               &  $-$  &   $-$   & $-$         \\  
Covering factor                                           & $-$                                   & $-$                                   & $0.5^{+0.4}_{-0.5}$                 & 0.1 (f)                               \\
\noalign{\vskip 0.5mm} \hline \noalign{\vskip 0.5mm}
$F_{\rm 2-10}$ \fluxcgs                    & $1.8\times10^{-13}$                   & $1.8\times10^{-13}$                   & $1.8\times10^{-13}$                   & $1.8\times10^{-13}$                   \\
%$F_{\rm 10-40}$ \fluxcgs                   & $1.3\times10^{-12}$                   & $1.3\times10^{-12}$                   & $1.3\times10^{-12}$                   & $1.3\times10^{-12}$                   \\
$L^{\rm int}_{\rm 2-10}$ \lumcgs           & $-$                                   & $1.6^{+0.6}_{-0.7}\times10^{43}$                    & $3.1^{+1.3}_{-1.4}\times10^{42}$                    & $1.1^{+0.5}_{-0.5}\times10^{43}$      \\
%$L^{\rm int}_{\rm 10-40}$ \lumcgs          & $-$                                   & $8.4\times10^{42}$                    & $3.0\times10^{42}$                    & $1.3\times10^{43}$                   
\enddata
\tablecomments{The intrinsic luminosity in the 2-10 keV band is computed from the intrinsic 2-10 keV flux, using the angular diameter distance from \citet{gao16}. Its uncertainty is estimated with the method described in the text and in Figure \ref{fig:multiple_contours}. For both MYTorus and Borus, the parameter $\Gamma$ is defined in the range [1.4-2.6], while $\log(N_{\rm H}/{\rm cm}^{-2})$ in the range [24-25] and [24-25.5], respectively. Hence, the symbols $+u$ and $-l$ mean that the parameter is capped to the upper and lower defined value, respectively.}
\end{deluxetable*}

\section{Intrinsic X-ray luminosity of the megamaser}\label{sec:lum}
\begin{figure*}
\plottwo{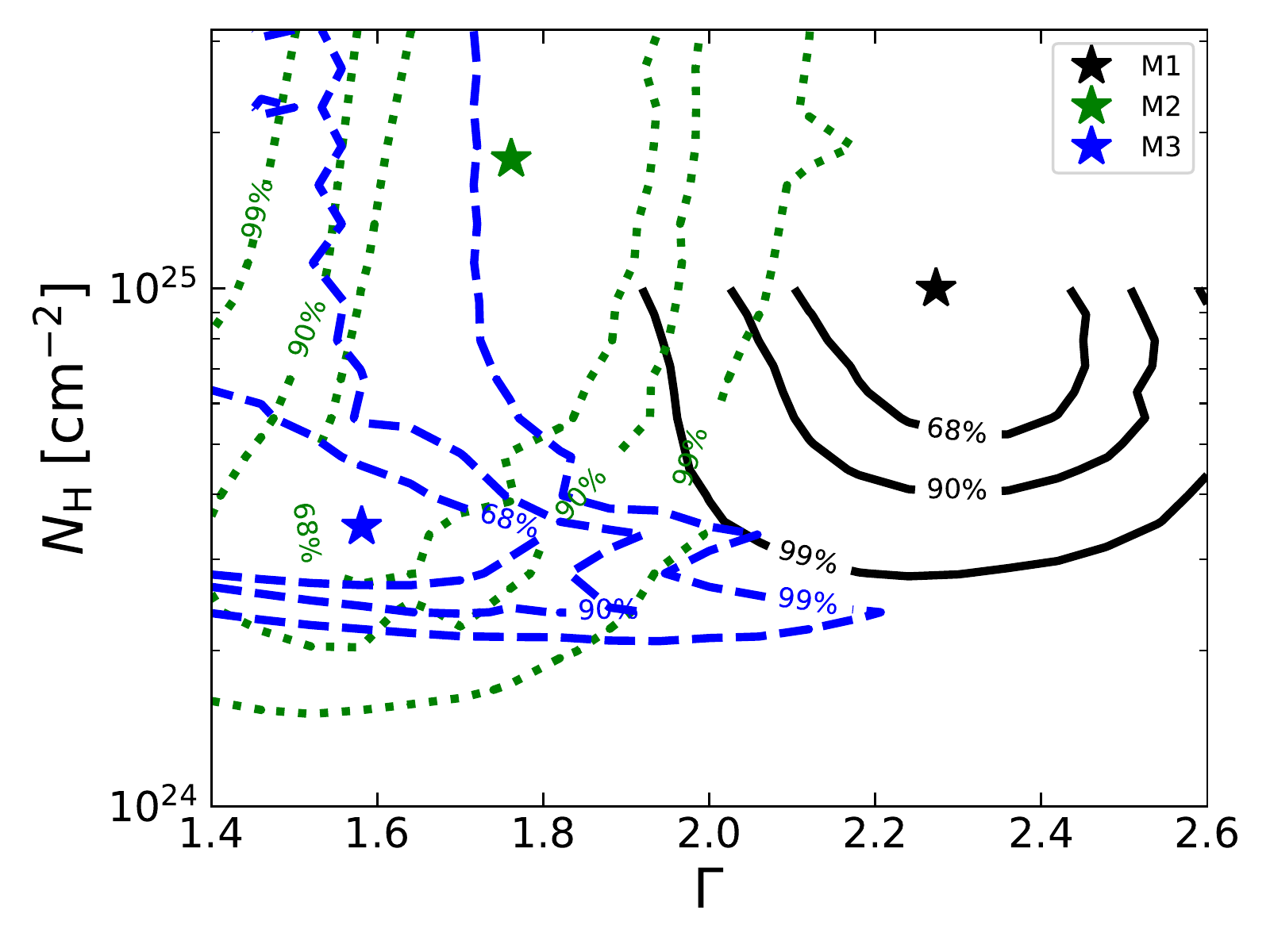}{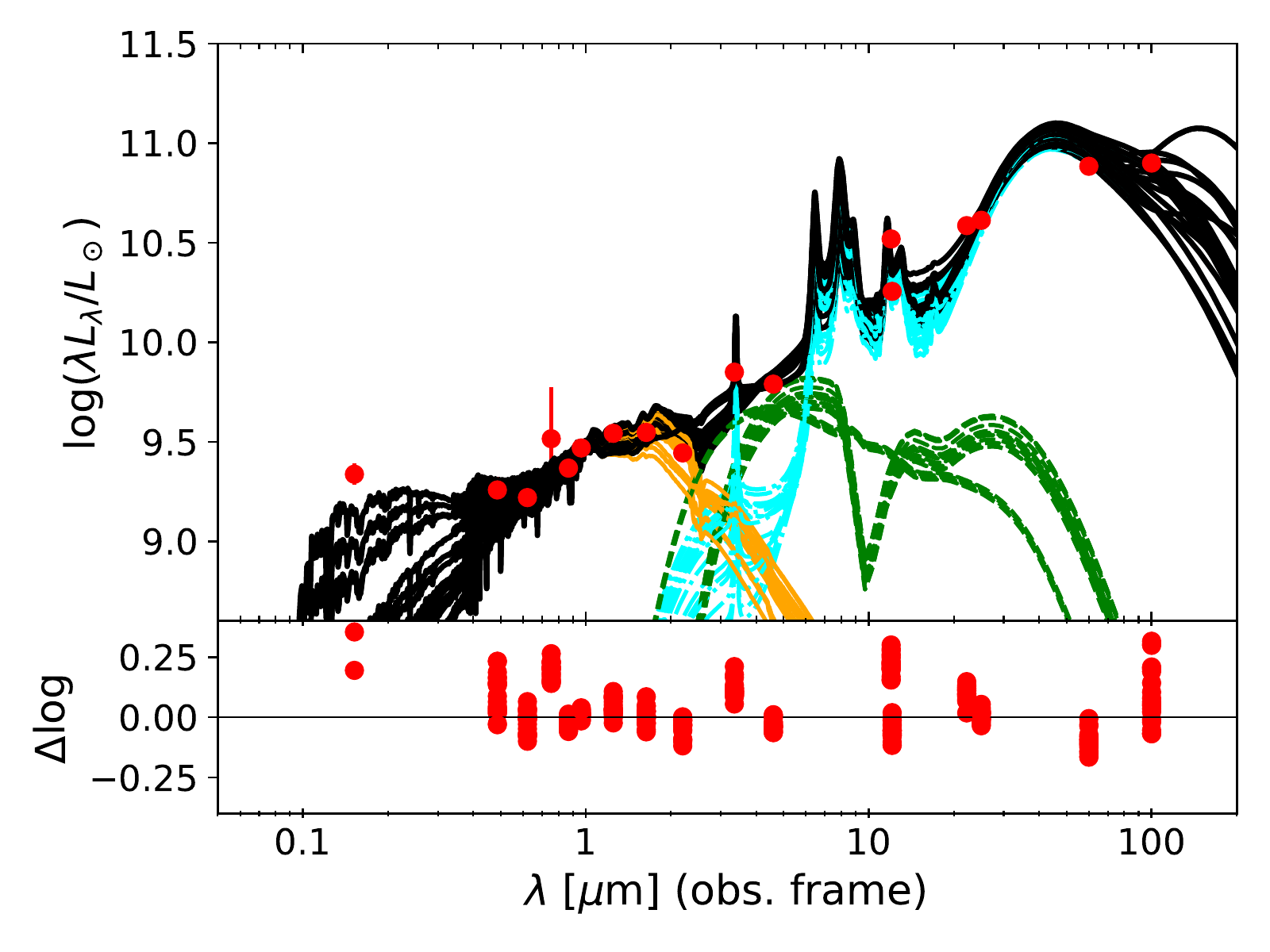}
\caption{\textbf{Left.} $\Gamma-$\nh confidence contour plot for the MYTorus model (solid black), Borus02 model in its general configuration (dotted green), and Borus02 model with a disk configuration (dashed blue). Contours are at the $68\%$, $90\%$ and $99\%$ of confidence limit, while the best fit values for the parameters are marked by colored stars. The \nh parameter caps at $10^{25}$ \nhcgs in the MYTorus model. While disagreeing on the intrinsic photon index, the models show that NGC 5765B is a bona-fide Compton thick AGN. The column density to which we are referring here is the one obscuring the line of sight. \textbf{Right.} SED fitting performed. All the 20 independent fits are shown, where the stellar component is labelled with an orange line, the cyan line labels star formation and the dashed green component is the AGN. The solid black line shows the total fit, and the red points are the data. The lower panel shows the deviation in $\Delta$log of the models from the data. \label{fig:multiple_contours}}
%Intrinsic $2-10$ keV luminosity as a function of $\Delta C$, for the different models explored in the spectral analysis: MYTorus (M1 - solid black line), Borus02 in its general configuration (M2 - dotted green line), and Borus02 in its disky configuration (M3 - dashed blue line). The two black dot-dashed horizontal lines mark the threshold for the $1\sigma$ and 90\% of confidence limit in the statistic.The colored vertical lines, along with the respective colored labels, mark the intrinsic $2-10$ keV luminosity expected from the 6 $\mu m$ \citep{lanzuisi09,fiore09,mateos15,stern15} or 12 $\mu m$ \citep{gandhi09,asmus15} luminosity, respectively. \label{fig:multiple_contours}}
\end{figure*}

In the previous Section, we have presented a comprehensive X-ray spectral analysis of the megamaser NGC 5765B. All the physically motivated models we employed (i.e., MYTorus and Borus02, models M1, M2, and M3) agree on the extreme obscuration affecting the intrinsic emission (see the left panel of Figure \ref{fig:multiple_contours}). This is supported also by the phenomenological model being reflection dominated (model M0). \par Model M0 could not provide an estimate of the intrinsic luminosity, not having a column density measurement and modeling a pure reflection dominated spectrum \citep{murphy09}. In any case, it is not straightforward to estimate the uncertainties on the luminosity directly from the spectral fitting. To this aim, we follow the methodology presented in \citet{boorman16}. We focus on the toroidal models (M1, M2, M3) and employ the uncertainties on the photon index and on the normalization of the primary continuum to build a grid of models, every one with its CSTAT. The intrinsic 2-10 keV luminosity is computed for every model in the grid. At a given luminosity, we choose the model with the lowest CSTAT in order to estimate the best fit luminosities and their $1\sigma$ uncertainties, which are reported in the last row of Table \ref{tab:results}. It can be seen that models M1 and M3  generally agree on the intrinsic 2-10 keV luminosity, while model M2 estimates a luminosity a factor of $\sim4-5$ lower than the other two. This suggests that its assumed geometry, a simple sphere with biconical cutouts, despite fitting the spectrum well, does not adequately describe the intrinsic properties of NGC 5765B. Indeed, neither the simple donut-shaped medium assumed by the default MYTorus model does.
 %On the other hand, since we decoupled the inclination of the different components of MYTorus, and indirectly assumed that the torus is clumpy (model M1), there is no reason why the line-of-sight \nh and the average column density of the torus (the \nh parameter of the Borus02 component) would need to be the same, leading to the definition of model M3. 
In model M3, the reprocessed spectrum predominantly comes from a geometrically thin and dense torus in the equatorial plane, but our line of sight intercepts a lower column density region, though still CT. Both models M1 and M3 require some decoupling of their fundamental components, supporting a non-uniform but rather clumpy obscuring medium.
\par To have an alternative (and complementary) measurement of the AGN luminosity, we collected available photometry from the NUV to the FIR bands, in order to decompose the spectral energy distribution (SED) of the megamaser. The adopted photometry is described in Table \ref{tab:sed}. We used \texttt{SED3FIT} \citep{berta13}, which is based on \texttt{MAGPHYS} \citep{dacunha08} and includes an AGN component based on the templates of \citet{fritz06} and \citet{feltre12}, to decompose the stellar, AGN and dust contributions. Based on the SED fitting, the AGN component dominates between 3-7$\mu m$, but one fit is unable to give reliable uncertainties on the bolometric luminosity of the AGN, $L_{\rm bol}$. Thus, we run 20 independent realizations (since \texttt{SED3FIT} uses random sampling of the libraries in order to speed up the process), which are shown in the right panel of Figure \ref{fig:multiple_contours}. The AGN component shows little spread throughout the fits, thus we take the mean and standard error on the mean as our bolometric luminosity with $1\sigma$ uncertainty ($L_{\rm bol} =2.75^{+0.07}_{-0.06} \times 10^{44}$ erg s$^{-1}$).

\begin{table}
\caption{Photometry adopted to fit the SED of NGC 5765B.}             
\label{tab:sed}      
\centering          
\begin{tabular}{l c} 
\hline\hline       
\noalign{\vskip 0.5mm} 
 Telescope/Band & Flux [Jy] \\ \noalign{\vskip 1mm} \hline  \noalign{\vskip 1mm}    
Galex/FUV &  $2.4\pm 0.3\times10^{-4}$  \\ \noalign{\vskip 0.5mm}
PanSTARRS/g & $6.5 \pm 0.9\times10^{-4}$   \\ \noalign{\vskip 0.5mm}
PanSTARRS/r & $7.57 \pm 0.08\times10^{-4}$   \\ \noalign{\vskip 0.5mm}
PanSTARRS/i & $1.8 \pm 0.8\times10^{-3}$   \\ \noalign{\vskip 0.5mm}
PanSTARRS/z & $1.49 \pm 0.08 \times10^{-3}$   \\ \noalign{\vskip 0.5mm}
PanSTARRS/Y & $2.1 \pm 0.1\times10^{-3}$  \\ \noalign{\vskip 0.5mm}
UKIDSS/J & $3.20 \pm 0.01\times10^{-3}$  \\ \noalign{\vskip 0.5mm}
UKIDSS/H & $4.23 \pm 0.01 \times10^{-3}$    \\ \noalign{\vskip 0.5mm}
UKIDSS/K & $4.49 \pm 0.01\times10^{-3}$    \\ \noalign{\vskip 0.5mm}
WISE/W1 & $1.74 \pm 0.04\times10^{-2}$  \\ \noalign{\vskip 0.5mm}
WISE/W2 & $2.08 \pm 0.04 \times10^{-2}$   \\ \noalign{\vskip 0.5mm}
WISE/W3 & $0.160 \pm 0.002$  \\ \noalign{\vskip 0.5mm}
WISE/W4  & $0.63 \pm   0.01$  \\ \noalign{\vskip 0.5mm}
IRAS/12$\mu m$*  & $0.29 \pm   0.03$  \\ \noalign{\vskip 0.5mm}
IRAS/25$\mu m$*  & $0.75  \pm  0.08$ \\ \noalign{\vskip 0.5mm}
IRAS/60$\mu m$*  & $3.4 \pm   0.3$  \\ \noalign{\vskip 0.5mm}
IRAS/100$\mu m$*  & $5.8 \pm   0.6$ \\ \noalign{\vskip 0.5mm}
\noalign{\vskip 1mm}    
\hline
\end{tabular}
\tablecomments{*We assumed a $10\%$ uncertainty on the IRAS fluxes. }
\end{table}

\subsection{The Eddington ratio of the megamaser}\label{sec:eddratio}

\begin{figure*}
\plotone{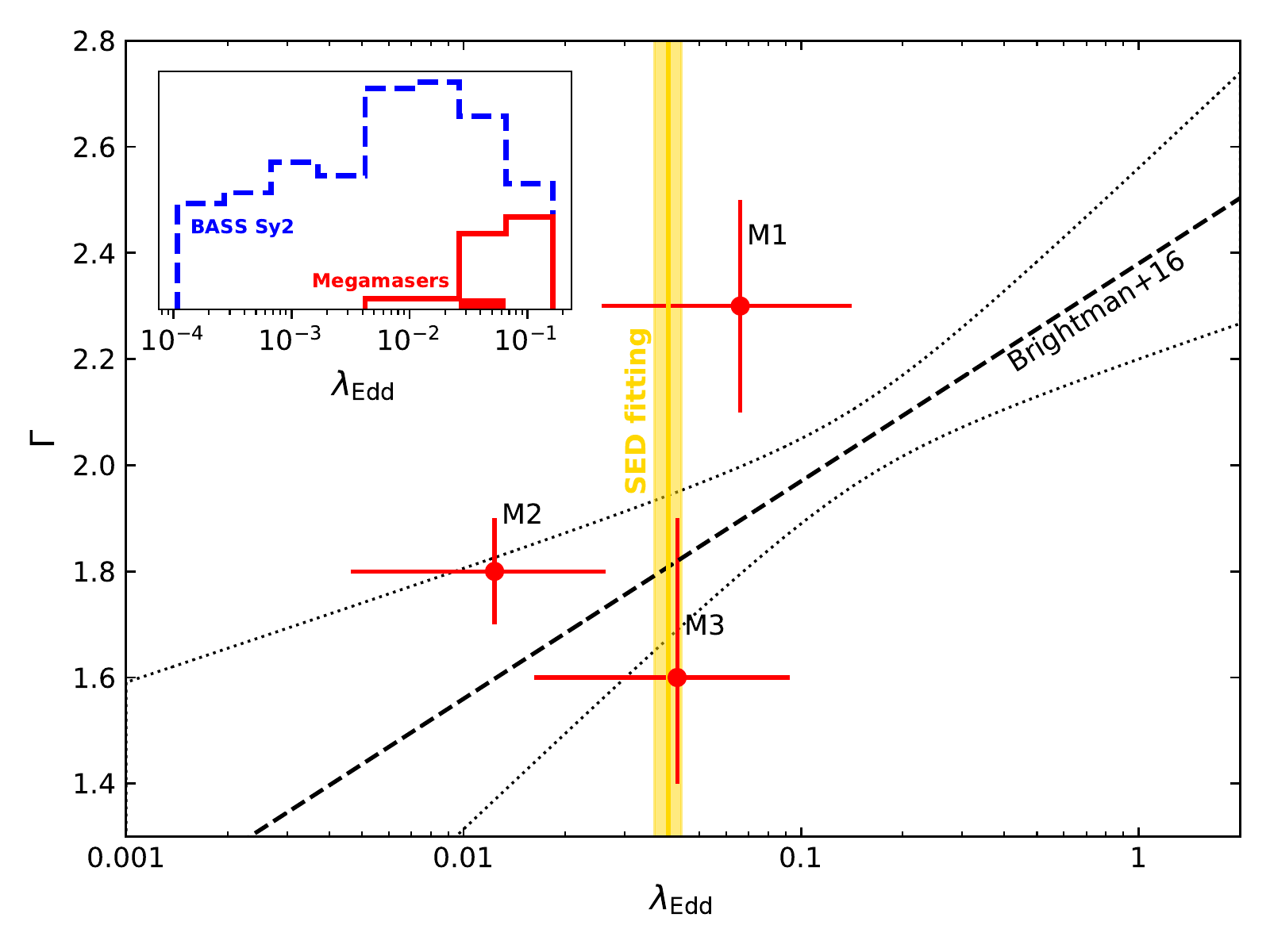}
\caption{Photon index - Eddington ratio plane, where the X-ray spectroscopic results for NGC 5765B adopting the three models discussed in the text are labeled with red points and error bars. The black dashed and dotted lines represent the best-fit relation found by \citet{brightman16} for megamasers and its relative uncertainty. The orange vertical band labels the Eddington ratio derived from SED fitting, and its uncertainty. The inset panel shows how the distribution of Eddington ratio for megamasers (red histogram; NGC 5765B is labeled in its bin in solid red) compares to the local \swiftbat Seyfert 2 AGNs from the BASS survey \citep[dashed blue histogram;][]{koss17}.\label{fig:gamma_eddrat}}
\end{figure*}

Adopting the  black hole mass $M_{\rm BH} = 4.55 \pm 0.40 \times 10^7 M_\odot$ \citep{gao16} and the bolometric correction factor $\kappa_{2-10} = 28^{+30}_{-14}$, suitable for CT AGN \citep{brightman17}, the Eddington ratio of NGC 5765B, derived from X-ray spectroscopy, is:

\begin{equation}
\lambda_{\rm Edd} = \begin{cases} 0.066^{+0.075}_{- 0.040}, & \mbox{model M1} \\ 0.043^{+0.050}_{- 0.027}, & \mbox{model M3},  \end{cases}
\end{equation}
%\\ 0.19^{+0.35}_{-0.15}, & \mbox{$L_{\rm X, IR}$} 
where the uncertainties are at $1\sigma$ confidence level. The uncertainty on $\lambda_{\rm Edd}$ is driven by the bolometric correction one. Using the bolometric luminosity from SED fitting, we obtain $\lambda_{\rm Edd} = 0.040\pm 0.004$, remarkably consistent with the Eddington ratios derived from the X-ray spectral analysis (in particular, the one of model M3) and broadly consistent with the one presented in \citet{kuo18}, estimated from the [OIII] luminosity from \citet{gao17}.
We note that model M3, our chosen best-fit model, provides also the most consistent pair of ($\Gamma$,$\lambda_{\rm Edd}$) with respect to the relation found by \citet{brightman16} using a sample of disk megamasers (see Figure \ref{fig:gamma_eddrat}). Disk megamasers usually display rather low-to-moderate Eddington ratio accretion \citep[$0.007 \lesssim \lambda_{\rm Edd} \lesssim 0.3$,][]{brightman16}, although they seem to place at the high-end of the Eddington ratio distribution of local, \swiftbat Sy2 AGNs \citep[see inset panel in Figure \ref{fig:gamma_eddrat}, and][]{koss17}; NGC 5765B, despite being in a merging stage, is no exception.
\par Taken at face value, such a low Eddington ratio could seem somewhat surprising given that mergers are expected to increase and facilitate nuclear accretion. However, the early stage of the merger, the potential presence of substantial time lags, and the intrinsic flickering nature of the accretion rate during AGN activity \citep{hickox14,schawinski15} prevent any strong link to be assessed between the two phenomena \citep{brightman18}. 
We can speculate that the AGN in NGC 5765B may have been triggered either by gravitational instabilities due to the potential of the companion galaxy, or that the AGN was already "on" before the two galaxies started merging. The diffuse tidal streams that can be seen in optical images and the strong kinematically disturbed HI emission detected in a VLA observation of the system \citep{pesce18} suggest that the two galaxies are now after their first encounter; hence, tidal torques may have triggered the AGN, as expected also from numerical simulations \citep{dimatteo05, capelo15}.

\section{Conclusions}\label{sec:conclusions}
In this paper, we have presented for the first time the hard X-ray spectrum of NGC 5765B, a disk water megamaser AGN in an early merging stage with a companion galaxy, hosting an AGN as well. Through a broadband spectral analysis, we confirmed that NGC 5765B is a bona fide, reflection-dominated CT AGN, and even the most up-to-date models fail to constrain the upper limit of its column density. Moreover, default configurations of such models are not able to reproduce its complex spectral shape, requiring to model a clumpy composition for the obscuring medium.  After correcting for such extreme obscuration, we estimated the Eddington ratio of the megamaser and found it to be in the range of few percent, consistent with the general trend of low-to-moderate accretion rate for disk water megamasers, and confirmed the result through SED decomposition. This is, to our knowledge, the most robust Eddington ratio ever derived for a merging galaxy. We argue that the nuclear activity of NGC 5765B is not showing any obvious and direct link to the ongoing merger. Future simulations, able to resolve gas flows on the smallest scales due to gravitational torques inducted by the first encounter in a merger, will provide the timescales associated to such phenomena and provide more clues on the role of mergers in triggering instabilities and AGN activity in the early phases of galaxy mergers.

%\par Given the early stage of the merger, the presence of a megamaser disk in a Keplerian rotation around the AGN on the sub-pc scale, the low-to-moderate Eddington ratio, and the general circumnuclear structure, with presence of both large scale and small scale spiral structrures \citep{pjanka17}, 

\acknowledgments
We thank the anonymous referee for useful suggestions that improved the clarity of the paper, and M. Balokovi\'c for useful help and suggestions on Borus02 modeling. 
\par This work was supported under NASA Contract NNG08FD60C, and made use of data from the \nustar mission, a project led by the California Institute of Technology, managed by the Jet Propulsion Laboratory, and funded by the National Aeronautics and Space Administration. We thank the \nustar Operations, Software, and Calibration teams for support with the execution and analysis of these observations. This research made use of the \nustar Data Analysis Software (NuSTARDAS) jointly developed by the ASI Science Data Center (ASDC, Italy) and the California Institute of Technology (USA). This research has also made use of data obtained from the \chandra\ Data Archive and the \chandra\ Source Catalog, and software provided by the \chandra\ X-ray Center (CXC).
A.M. and R. C. H. acknowledge support by the NSF through grant numbers 1554584, and by NASA through grant numbers NNX15AP24G and Chandra GO award GO7-18130X.
M.K. acknowledges support from NASA through ADAP award NNH16CT03C.
A.C., M.B and G.L. acknowledge support from the ASI/INAF grant I/037/12/0-011/13.

\vspace{5mm}
\facilities{\chandra, \nustar}

\software{CIAO \citep{fruscione06}, NuSTARDAS, XSPEC \citep{arnaud96}}
\bibliographystyle{aasjournal} % style aasjournal.bst 
%\bibliography{/media/alberto/TOSHIBA2/amasini} % your references Yourfile.bib
\bibliography{/Volumes/Maxtor/amasini} % your references Yourfile.bib

\end{document}